\documentclass[twocolumn,nofootinbib,longbibliography,superscriptaddress]{revtex4-1}
\usepackage{graphicx}
\usepackage{dcolumn}
\usepackage{bm,bbm}
\usepackage{xcolor}
\usepackage{tcolorbox}
\usepackage{algorithm}
\usepackage{algpseudocode}
\usepackage{amsmath}
\usepackage{bbold}
\usepackage{braket}
\usepackage{nameref}
\usepackage[toc,page]{appendix}

\usepackage[colorlinks,breaklinks,linkcolor={blue},citecolor={magenta},urlcolor={blue}]{hyperref}
\usepackage{enumerate}

\usepackage{multirow}

\usepackage[commandnameprefix=always]{changes}

\usepackage{ulem}
\usepackage{xcolor}

\begin{document}

\title{Topological Quantum Interferometry}

\author{Tianyou Ying}
\thanks{These authors contributed equally to this work.}
\affiliation{National Laboratory of Solid State Microstructures, School of Physics, and Collaborative Innovation Center of Advanced Microstructures, Nanjing University, Nanjing 210093, China }

\author{Yufeng Zhou}
\thanks{These authors contributed equally to this work.}
\affiliation{National Laboratory of Solid State Microstructures, School of Physics, and Collaborative Innovation Center of Advanced Microstructures, Nanjing University, Nanjing 210093, China }

\author{Chengwei Pan}
\affiliation{National Laboratory of Solid State Microstructures, School of Physics, and Collaborative Innovation Center of Advanced Microstructures, Nanjing University, Nanjing 210093, China }

\author{Ryan Hogan}
\email{ryan.hogan@nju.edu.cn}
\affiliation{National Laboratory of Solid State Microstructures, School of Physics, and Collaborative Innovation Center of Advanced Microstructures, Nanjing University, Nanjing 210093, China }

\author{Ruoyang Zhang}
\affiliation{National Laboratory of Solid State Microstructures, School of Physics, and Collaborative Innovation Center of Advanced Microstructures, Nanjing University, Nanjing 210093, China }

\author{Hui Liu}
\email{liuhui@nju.edu.cn}
\affiliation{National Laboratory of Solid State Microstructures, School of Physics, and Collaborative Innovation Center of Advanced Microstructures, Nanjing University, Nanjing 210093, China }

\author{Shining Zhu}
\affiliation{National Laboratory of Solid State Microstructures, School of Physics, and Collaborative Innovation Center of Advanced Microstructures, Nanjing University, Nanjing 210093, China }

\author{Xiaoqin Gao}
\email{xiaoqin.gao@nju.edu.cn}
\affiliation{National Laboratory of Solid State Microstructures, School of Physics, and Collaborative Innovation Center of Advanced Microstructures, Nanjing University, Nanjing 210093, China }

\begin{abstract}
Structured light provides high-dimensional Hilbert spaces holding tremendous potential for fundamental quantum optics and quantum technologies. However, existing characterization methods, like Hong-Ou-Mandel (HOM) interference, typically assume perfectly tuned conditions, overlooking the geometric physics governing spatial mode evolution. Here, we establish topological quantum interferometry driven by an interaction-based geometric phase, the exchange Berry phase (BPX). Our formalism generalizes $q$-plate state generation and characterization to arbitrary topological charges and (de)tuning conditions, demonstrating that BPX acts as a geometric marker governing spatial interference. We show BPX serves as a deterministic control parameter, decomposing two-photon spatial patterns into geometry-dictated fundamental modes. This mapping reveals topological invariants and phase singularities that function as a non-tomographic witness for state dimensionality estimation, circumventing full-state reconstruction. Being device-independent and highly scalable, this approach enables scalable high-dimensional characterization and topologically protected state selection, with direct applicability to quantum metrology and high-capacity quantum networks.

\end{abstract}

\maketitle

\section*{Introduction}
Hong-Ou-Mandel (HOM) interference is a fundamental tool for characterizing bipartite systems and photon indistinguishability~\cite{hong1987measurement, bouchard2021two}. While traditionally used for light source characterization~\cite{chapman2025chip}, quantum computing~\cite{o2003demonstration, kok2007linear}, Boson sampling~\cite{zhong2021phase,zhong2020quantum}, optimal quantum cloning~\cite{nagali2009optimal,irvine2004optimal,bouchard2017high},  quantum communication~\cite{lo2012measurement}, and {quantum} metrology~\cite{lyons2018attosecond,harnchaiwat2020tracking}, its application with high-dimensional states offers a route to significantly enhance the information capacity of quantum protocols and quantum information processing. Utilizing the spatial degree of freedom is an effective route toward higher dimensionality, which inherently increases capacity.

Devices like $q$-plates are versatile for this purpose, as they couple spin to orbital angular momentum (OAM), enabling arbitrary conversion between polarization and spatial modes~\cite{marrucci2006optical}. These devices are widely used to generate vector modes (VMs)~\cite{cardano2012polarization,slussarenko2011tunable,d2015arbitrary}, which exhibit spatially varying polarization structures arising from geometric phase, known as Berry phase~\cite{berry1987adiabatic}. Berry phase has been well studied in optics and condensed matter, playing a vital role in understanding the geometry and topology~\cite{whitney2005geometric,milione2012higher,leinonen2023noncyclic}. {While typically considered Abelian, that is, order of operations is irrelevant, Berry phase has also been shown to be non-Abelian~\cite{cheng2025non}, which is highly relevant to quantum computing~\cite{pachos1999non}. In a system of complex operations, especially in the case of partial spin-to-OAM conversion, non-Abelian physics could be possible. Moreover, quantum state evolution associated with Berry phase is possible when considering parametrization using synthetic dimensions~\cite{yang2024non}, and is highly relevant to topological photonics,~\cite{ lustig2021topological,ehrhardt2023perspective}.}

For photonic systems, Berry phase is not measurable from single-photon intensities. However, considering two-photon interference, like HOM interferometry,  spatial phase information within the interference terms can produce complex spatial structures, which can be represented by an interaction of each photon's geometric phase. It is precisely this phase we wish to study as the marker for the underlying physics, which we deem the exchange Berry phase (BPX). Overall, the resulting spatial pattern properties have led to broad applications in quantum key distribution~\cite{vallone2014free,souza2008quantum}, high-capacity data transmission~\cite{bozinovic2013terabit}, classical and quantum communication~\cite{d2012complete,farias2015resilience,li2016high,ndagano2017creation}, quantum microscopy~\cite{yan2015q}, high-resolution imaging~\cite{chen2007numerical}, and fundamental experiments~\cite{karimi2014hardy,cardano2013violation}. Therefore, research regarding the foundation of the underlying physics and how Berry phase and BPX govern photon interaction must be developed.

\begin{figure*}[t!]
\includegraphics [width=\textwidth]{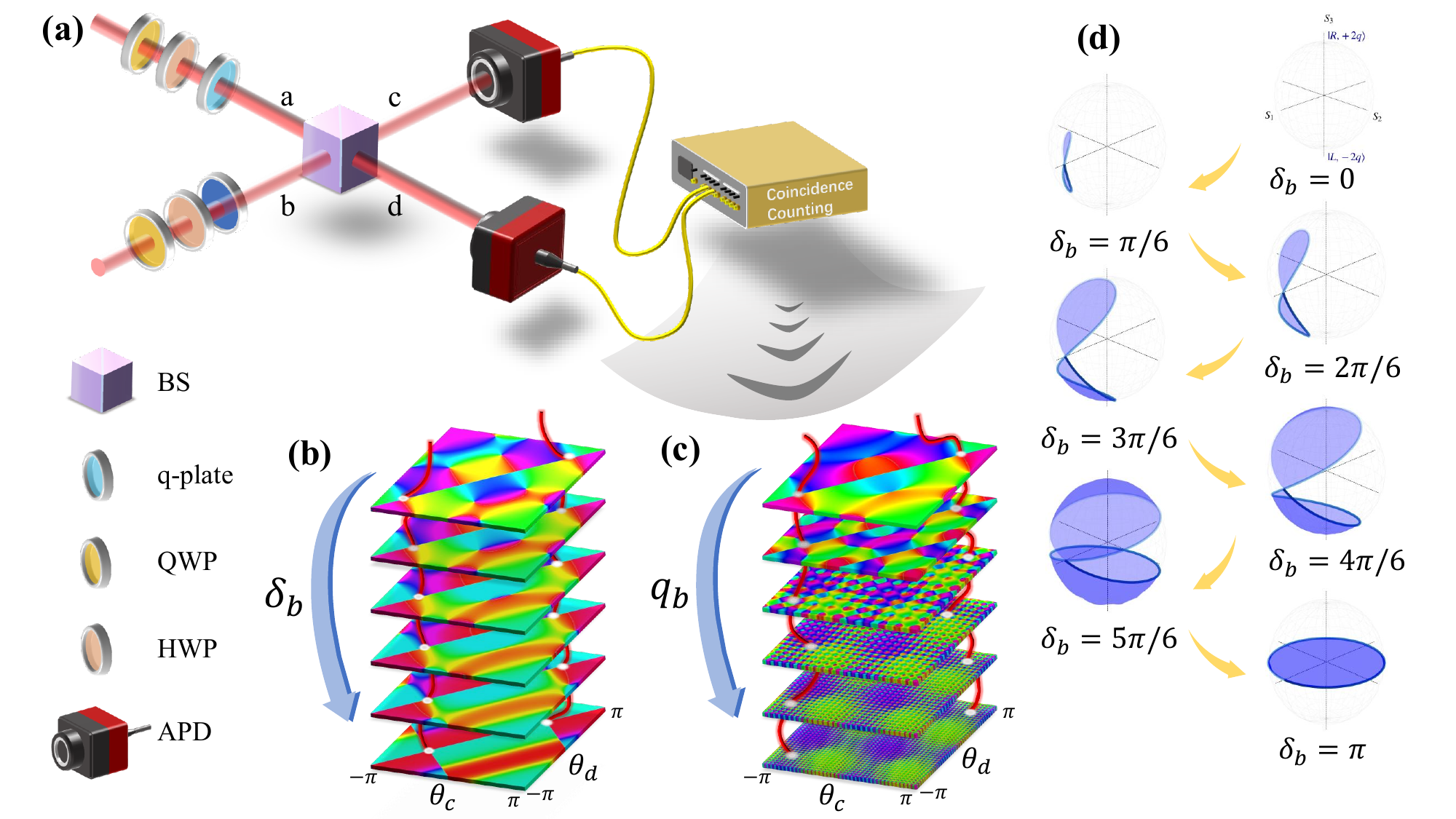}
\centering
\caption{\textbf{Conceptual framework of topological quantum interferometry.} \textbf{(a)} Two structured photons, prepared using quarter-wave plates (QWPs), half-wave plates (HWPs), and independent $q$-plates ($q_a, \delta_a$ and $q_b, \delta_b$) in input paths $a$ and $b$, respectively, interact at a 50:50 beamsplitter (BS) and undergo Hong-Ou-Mandel (HOM) interference. The spatial dependence of the resulting Bell-state probabilities and interference visibility is monitored in coincidence between avalanche photodiodes (APDs) in the output paths $c$ and $d$. The overall pattern is dictated by the exchange Berry phase (BPX), which represents an interaction geometric phase arising from the constituent photons' topologies. \textbf{(b)} Schematic of the evolution of phase of the Bell States representing the BPX for increasing birefringence detuning, $\delta_b$, with $\delta_a=\pi/2$, $q_a=1$ and $q_b=-0.5$. Here, the phase singularities (white dots) carry topological invariants (i.e., the phase winding number around them) at each detuning from $0$ to $\pi$, which spatially represent noise-robust conditions. Red strings indicate the movement, or lack thereof, of said phase singularities as parameters are tuned, demonstrating their persistence \textbf{(c)}. Similarly, the evolution for various topological charges, $q_b$, is shown for $q_b=[0.5,2,5,10,15,20]$, where spatial frequency increases with large $q_b$. Note that here, $q_a=1$, $\delta_a=\pi/2$, and $\delta_b=\pi/2$. Here, 
pathways persist passing through phase singularities. However, we note that an increase in $q_b$ leads to a higher density of phase singularities, and therefore, more possible pathways overall. We observe in \textbf{b} and \textbf{c} that BPX serves as a deterministic control parameter for Bell state probability evolution and interference visibility. Monitoring BPX enables the identification of state dimensionality (here as OAM) without full-state tomography.  \textbf{(d)} The evolution for increasing $\delta_b$ of a single-photon (path $b$) quantum state on the Poincaré sphere, with axes defined by Stokes Parameters, $S_i$. Initially, the beam is Gaussian, and there is no observed azimuthal coordinate trajectory. Different radii modify the trajectory; however, we assume a uniform radial distribution (more details provided in the Supplementary Material). As the spin (here, polarization) to orbital angular momentum (OAM) conversion increases, the trajectory progressively enlarges until it is a circular path around the equator for a fully tuned q-plate ($\delta_b=\pi$).}
\label{fig:schematic}
\end{figure*}

One method is examining spatial correlations, where we can gain deeper insight into the spatial effects within quantum interference. These insights are highly applicable to spatially varying polarization and complex field structures, which can uniquely be used to generate four Bell states and can be well characterized~\cite{d2016entangled,gao2025generation,gao2024full,schiano2026tailoring}. Current demonstrations, however, are typically constrained to perfectly tuned states ($\delta = \pi$). Partially tuned $q$-plates ($\delta \neq \pi$) generate complex coherent superpositions that significantly expand the available Hilbert space, but a detailed theoretical analysis of two-photon interference under {all} conditions has not been established. A universal model describing the dynamic evolution under all conditions would offer critical insight into the geometric nature and generation mechanisms of vector-vortex fields.

In this letter, we introduce a comprehensive theoretical model for {topological quantum interferometry with (de)tuned $q$-plates.} 
We analyze the spatial probability distributions of Bell states and their associated visibility across {various (de)tunings} and topological charges, and link the concept of
BPX, which allows direct inference of the evolution of Bell state probabilities and interference visibility. BPX toggling is scaled by multiplicative detuning terms, which affect the conversion amplitude of polarization to spatial modes. Furthermore, we identify phase singularities with topological invariants (i.e. winding numbers) that persist across different detunings and charges. Such singularities in BPX provide a non-tomographic witness to estimate state dimensionality without full state tomography. Furthermore, we briefly discuss how BPX could be linked to non-Abelian behaviour in scenarios of complex operation, which could pave the route toward holonomic quantum computing~\cite{pachos1999non}. Overall, our approach
enables {a device-independent, high-precision characterization, and deterministic selection} of complex structured photon states, unlocking new possibilities for high-dimensional quantum information processing.

\section*{Concept}
We begin by examining the action of a $q$-plate on circularly polarized light. For a given topological charge $q$ and birefringence detuning $\delta$, the $2\times2$ unitary transformation matrix, $\hat{U}_{q\mathrm{\text{-}plate}}$, is
\begin{equation}
\label{eq:1}
    \hat{U}_{q\mathrm{\text{-}plate}} = \hat{R}(-\gamma) \hat{D}(\delta) \hat{R}(\gamma)
\end{equation}
where $\hat{R}(\pm\gamma)$ is a rotation matrix, and the argument is $\gamma=2q\theta$. Here, $\theta$ represents the azimuthal coordinate, and $\hat{D}(\delta)$ is the detuning phase matrix (For details, see the Supplementary Material). Here, $\hat{U}_{q\mathrm{\text{-}plate}}$ acts on the polarization basis $\{\ket{L},\ket{R}\}$ as defined in Ref.~\cite{gao2025generation}. The radial dependence of hypergeometric functions~\cite{allen1992orbital}, $F_{q}(r)$, associated with Laguerre-Gaussian modes ($p=0, l=2q, z=0$) are also discussed in the Supplementary Material.

Consider two input beams in paths $a$ and $b$ that independently pass through a series of quarter-wave plates (QWPs), half-wave plates (HWPs), and $q$-plates, each with their unique detunings and topological charges ($\delta_a, q_a$ and $\delta_b, q_b$). These beams then interact at a 50:50 beam-splitter (BS), and are analyzed in coincidence using two avalanche photodiodes (APDs) in paths $c$ and $d$, as shown in Fig. \ref{fig:schematic}(a). We observe the antibunching scenario, from which we decompose the output state, $\ket{\Psi}$, into the Bell basis $\{\ket{\psi^\pm}, \ket{\phi^\pm}\}$ 
\begin{equation}
\ket{\Psi} = \sum_{\mathcal{B}} c_{\mathcal{B}} \ket{\mathcal{B}}, \quad \mathcal{B} \in {\psi^{\pm}, \phi^{\pm}},
\label{eq:psi}
\end{equation}
where the complex amplitudes $c_{\mathcal{B}}$ of each state $\mathcal{B}$ encapsulate the contribution of each Bell component. Here, the primary observables are the spatial patterns of Bell state probabilities and interference visibility, each highly sensitive to detuning and topological charge. The underlying physics governing such spatial behaviours follows the collective effect of constituent photons' Berry phase, which uniquely defines BPX. 

In Fig. \ref{fig:schematic}\textbf{(b)} and \textbf{(c)}, we show a schematic of {the evolution of the phase distribution of a singular Bell state ($\phi^+$), which is schematically representative of how BPX varies with respect to different detunings} and topological charges, respectively. We show here how topological invariants (depicted as red strings) arise in BPX by following phase singularities (depicted as white circles). Through these
pathways, the state can be considered noise-robust. Moreover, we show that the presence of phase singularities acts as a non-tomographic witness for state dimensionality estimation, which can significantly alleviate the computational costs associated with full state tomography. Together, the robustness to noise and non-tomographic witnesses are excellent to improve modern quantum protocols. {However, to understand BPX and all its utility, we must contextualize} the importance of the constituent photons' Berry phase and how that quantitatively gives rise to BPX. 

The constituent photon Berry phase is easily visualized as the geometric phase acquired upon smooth evolution of a {single-photon} quantum state on the Poincaré sphere, as shown in Fig. \ref{fig:schematic}\textbf{(d)}. Here, we note that the shape of the single-photon quantum state trajectory is independent of the topological charge, but the overall winding number changes around the singularities in the two-photon BPX distribution. It is also important to note that the overall conversion of spin angular momentum into OAM is governed by both $q$ and $\delta$. Interestingly, the Berry phase remains a non-observable global offset in the single-beam picture; however, the interference of two such beams manifests as the BPX, mapping the constituent geometric phases as an interaction phase onto the azimuthal plane. As a result, the two-photon correlation intensity and spatial patterning are affected and can be directly controlled. {In essence, the BPX represents the spatial pattern boundaries, and the detuning scales the BPX terms, resulting in overall amplitude conversion from Gaussian to vector-vortex beams. Here, this action sets the Bell state probability and interference visibility patterns.} We will come to see that $\delta$ strongly affects the visibility and amplitude, while $q$ affects the spatial frequency of the patterns, and the overall amount of phase singularities.

Consider the case when two beams with different topologies interact; their relative spatial geometry evolves, leading to spatially varying correlations and patterning distributed about azimuthal coordinates $\theta_c$ and $\theta_d$. {We uniquely define the BPX phase as
\begin{equation}
\Phi_\mathrm{BPX}^{\pm}(\theta_c, \theta_d) = 2(q_a\theta_c \pm q_b\theta_d),
\label{eq:bpx}
\end{equation}}
and $\Phi_{\mathrm{BPX}}'$ indicates a coordinate change such that $\theta_c\rightarrow\theta_d$ and vice versa. 
While the representation is for a two-photon interaction, the concept can be generalized and will be the subject of future studies.
For Gaussian beams ($q_a=q_b=0$), $\Phi^{\pm}_{\mathrm{BPX}}=0$, confirming that spatial patterns arise strictly from the topological geometry. A fully tuned ($\delta=\pi$) $q$-plate generates a complete spin-to-OAM conversion, while partially tuned cases result in a superposition of Gaussian and OAM modes. Detuning modifies the quantum state trajectory {of a single photon} on the Poincaré sphere, where, interestingly, small detunings (e.g., $\delta_b = \pi/6$) imprint complete geometric patterns, albeit with reduced contrast. As a result, HOM interference visibility is directly influenced.

\begin{widetext}
Accounting for partial tuning, these amplitudes split cleanly into orthogonal polarization channels via selection rules. For $\psi^{\pm}$ states, cross- terms vanish under projection, leaving
\begin{equation}
\mathcal{F}_{\psi^{\pm}} = \left[ C_a C_b - S_a S_b e^{-i\Phi_{\mathrm{BPX}}^-} \right] \pm \left[ -C_a C_b + S_a S_b e^{-i\Phi_{\mathrm{BPX}}'^-} \right]
\end{equation}

\begin{figure*}
\includegraphics [width=1\textwidth]{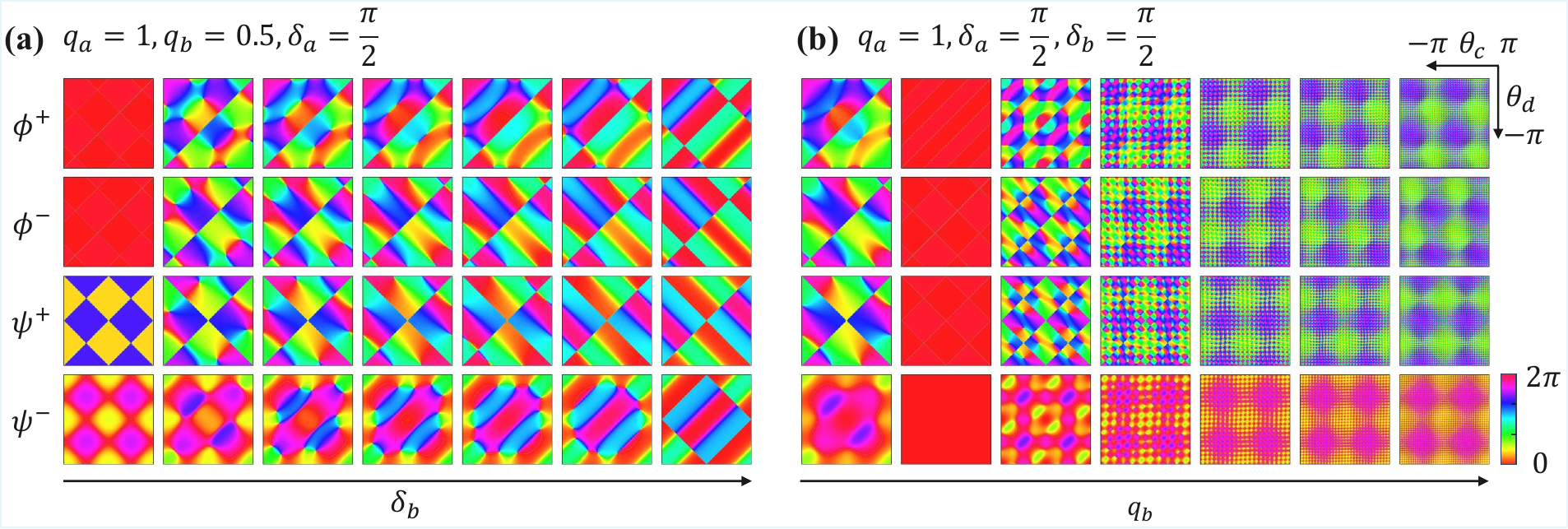}
\centering
\caption{\textbf{Evolution of the {phase of the Bell states representing the BPX} across the Bell basis.}  \textbf{(a)} Variation of the detuning parameter $\delta$ over the range $[0, \pi]$ demonstrates that topological structure remains invariant regardless of the amount of the converted spin-to-OAM. Here, the detuning modulates the amplitude of the spatial modes, where the observed transition directly governs the interference visibility. This transition is a quantitative equivalent of the state evolution depicted on the Poincaré sphere in Fig. \ref{fig:schematic}\textbf{d}, however,  states are now represented in the Bell basis $\{\ket{\psi^\pm}, \ket{\phi^\pm}\}$. \textbf{(b)} Variation of the topological charge $q_b$ for $q_b=[0.5,2,5,10,15,20]$ demonstrates scaling of the spatial frequency of the interference patterns by BPX while maintaining fundamental topological invariants. The azimuthal orientations of the fringes dictate the correlation (diagonal) and anti-correlation (mirrored) manifolds within the Bell state probability distributions.}
\label{fig:berryphase}
\end{figure*}

Conversely, $\phi^{\pm}$ states are written as
\begin{equation}
\begin{aligned}
\mathcal{F}_{\phi^{\pm}} &= \frac{i}{2}\Big(\left[ C_a S_b e^{-i2q_b\theta_d} + S_a C_b e^{i2q_a\theta_d} \right] \pm \left[ C_a S_b e^{-i2q_b\theta_c} + S_a C_b e^{i2q_a\theta_c} \right]\Big)\\
&= \frac{i}{2}\Big(e^{i\Phi_a} \left[ S_a C_b + C_a S_b e^{-i\Phi_{\mathrm{BPX}}^{+}}\right] \pm e^{i\Phi_a'} \left[ S_a C_b + C_a S_b e^{-i\Phi_{\mathrm{BPX}}'^{+}}\right]\Big),
\end{aligned}
\end{equation}
\end{widetext}
where $C_{a,b}=\cos(\delta_{a,b}/2)$, $S_{a,b}=\sin(\delta_{a,b}/2)$ and $\Phi_a=2q_a(\theta_c-\theta_d)$, with prime again indicating coordinate flip. This expression confirms that even for identical photons ($q_a = q_b$), the specific geometric terms reinforce or cancel at specific coordinates. The resulting coincidence probability is a map of the local intensity profile, $\mathcal{I}(\theta_c, \theta_d) \propto |\mathcal{F}|^2$, which governs the two-photon spatial correlations. Note that for identical detuning ($\delta_a = \delta_b = \delta$) in the $\psi^-$ channel, the visibility captures the topological separation of the constituents, modifying the two-photon output visibility:
\begin{equation}
\nu(\delta) = \frac{\sin^2(\delta/2)}{\cos^2(\delta/2) + 1}.
\end{equation}
At $\delta = \pi$, we obtain maximum visibility from where the geometric phases act as toggles. We can generate purely symmetric (antisymmetric) states when both $\Phi_{\mathrm{BPX}}^{-}$ and $\Phi_{\mathrm{BPX}}'^{-} $ simultaneously go to zero ($\pi$), albeit a stringent condition. By modulating the photons' exchange symmetry at the beam splitter, a geometric rotation can be applied to the symmetric components of the joint wavefunction; we can map them onto $\ket{\psi^-}$, thereby achieving maximal coincidence counts. In general, however, we note that the spatial structure of the interference follows $\nu = (C_{\mathrm{out}} - C_{\mathrm{in}}) / C_{\mathrm{out}}$ as defined across the $(\theta_c, \theta_d)$ coordinate space~\cite{schiano2026tailoring}.

The dimensionality $d$ of the Hilbert space representing the joint two-photon state can be estimated by the number of solutions to the equations $\mathcal{F}_{\mathcal{B}}(\theta_c, \theta_d) = 0$, with $\mathcal{B}$ defined by Eq.~\eqref{eq:psi}. For example, the case $q_a=1.0$ and $q_b=-0.5$ sets $e^{-i\Phi_{\mathrm{BPX}}^-}$ to wind through the coordinate space such that at least three distinct topological phase resets are required to satisfy the boundary conditions of the $2\pi \times 2\pi$ detector plane. From here, the net topological winding, ${W}$, can be described as follows
\begin{equation}
W = \frac{1}{2\pi} \oint_{\partial S} \vec{A}_{\mathrm{BPX}} \cdot d\vec{\ell},
\end{equation}
where $\vec{A}_{\mathrm{BPX}} = \vec{\nabla} \Phi_{\mathrm{BPX}}^\pm = \left( \frac{\partial \Phi_{\mathrm{BPX}}^\pm }{\partial \theta_c}, \frac{\partial \Phi_{\mathrm{BPX}}^\pm }{\partial \theta_d} \right)$ represents the effective geometric connection vector, and $\partial S$ denotes the closed boundary enclosing the $(\theta_c, \theta_d)$-coordinate space. This closed loop integral evaluates the net topological charge of the enclosed phase singularities, providing an approximate estimate of the Hilbert space dimension, $d$, where $W \approx d$.

For our system, the unitary operations commute; however, more complex systems could be non-commutative, and therefore non-Abelian.
Here, path-order sensitivity in quantum interference could be possible, providing a useful step towards wider utility in applications such as quantum computing~\cite{pachos1999non} and other non-Abelian topological photonics~\cite{zhang2022non,cheng2025non}. Furthermore, by transferring the macroscopic variables to synthetic dimensions~\cite{yang2024non,lustig2021topological,ehrhardt2023perspective}, such as topological charge, q, detuning, $\delta$, or BPX phase singularities, new avenues for fundamental and application-based studies of non-Abelian physics could be possible.

\section*{Results}
We now quantitatively analyze the spatial dependence of BPX, Bell state probabilities, and interference visibility under various detuning conditions and various topological charges. For two interacting structured beams with topological charges ($q_a,q_b$) and detunings ($\delta_a,\delta_b$) from input paths ($a,b$), spatially-dependent correlation patterns from output paths ($c,d$) are plotted in azimuthal coordinates $\theta_c$ and $\theta_d$. 

Figure \ref{fig:berryphase}\textbf{(a)} demonstrates the effect of detuning $\delta_b$ on {the phase of the Bell states, which corresponds to the BPX distribution,} with $q_a = 1, \delta_a=\pi/2, q_b=-0.5$. Progression from $0$ to $\pi$ induces a continuous change {in the contrast of fringes, representing the overall conversion of spin-to-OAM amplitude of the constituent photons as the Berry phase is mapped onto the photon spatial modes. Further analysis reveals $\ket{\psi^\pm}$ exhibits a relative $\pi$ phase shift with respect to each other. In contrast, the $\ket{\phi^\pm}$ involve anti-correlations in the spin-OAM basis. The resulting fringes are governed by single-photon cross-conversion phases localized to individual output ports (e.g., $2q_a\theta_d + 2q_b\theta_d$), creating an orthogonal spatial orientation compared to the diagonal structure of the $\ket{\psi^\pm}$. Otherwise stated, the azimuthal dependence undergoes a structural coordinate shift proportional to $\Phi_a$ or $\Phi_a'$.

\begin{figure}
\includegraphics [width=\columnwidth]{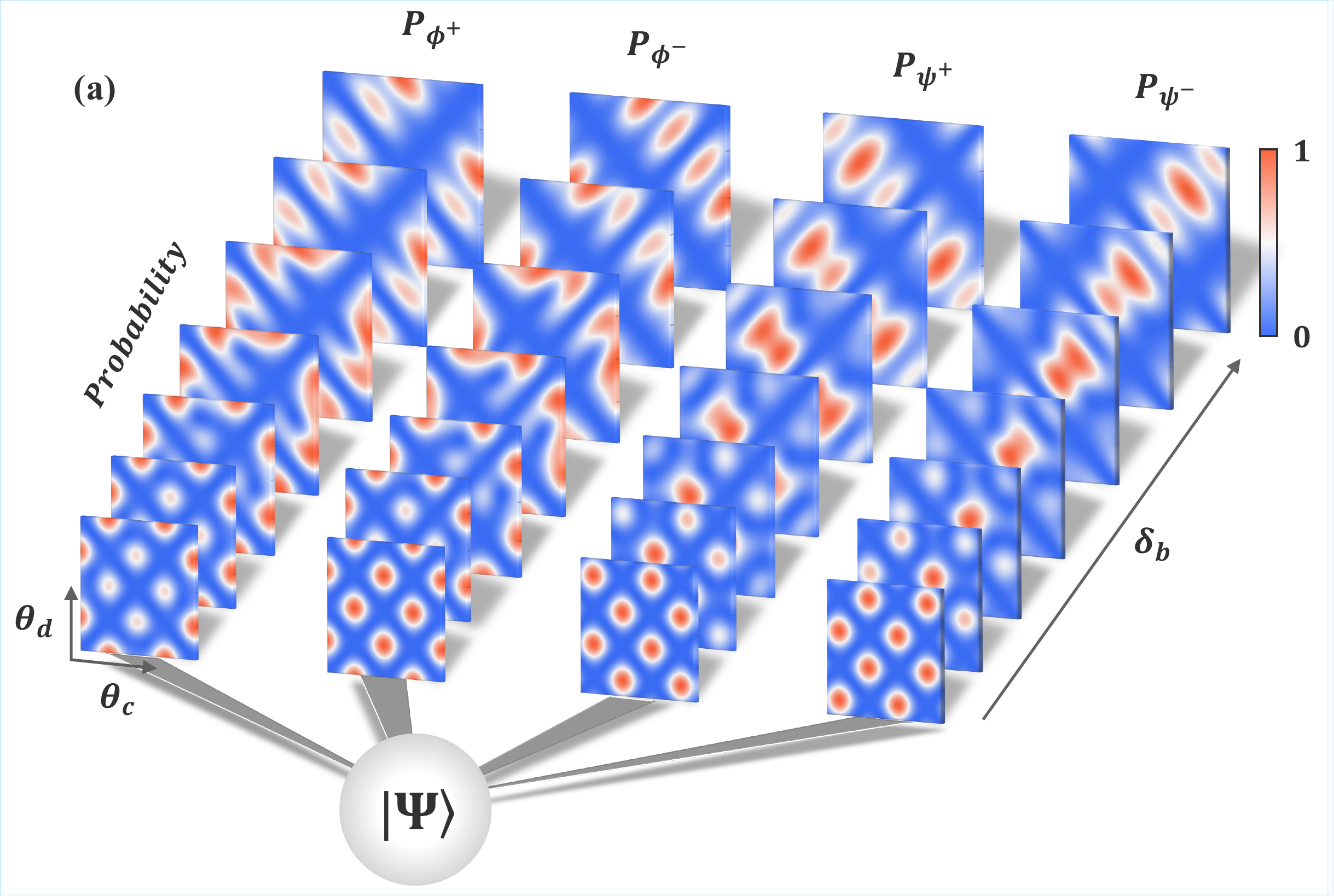}\\ [5mm]
\includegraphics [width=\columnwidth]{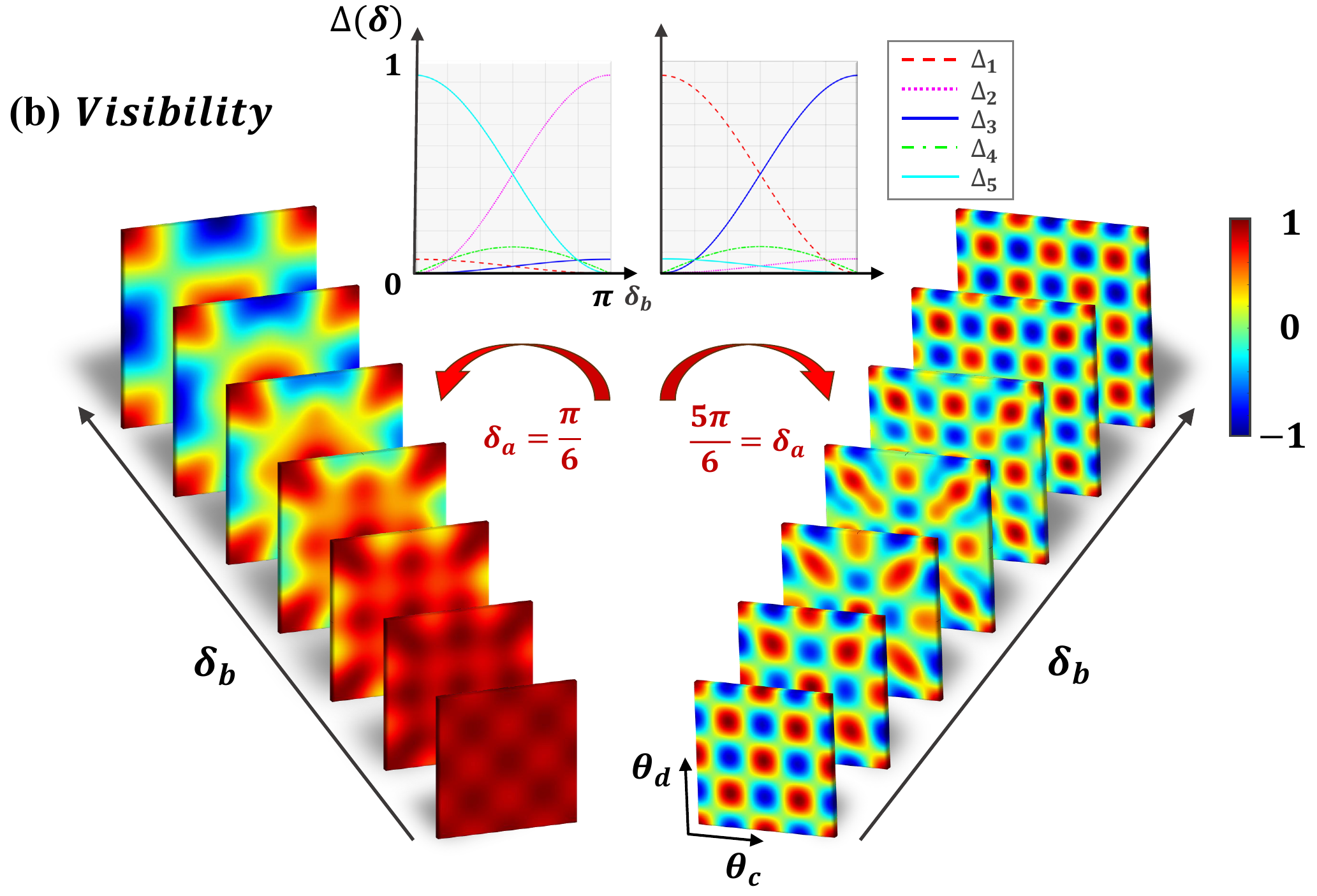}
\centering
\caption{\textbf{Deterministic control of spatial patterns under various detuning.} \textbf{(a)} Normalized spatial probability distributions of the four Bell states as a function of the detuning $\delta_b$ with $\delta_a = \pi/2$, $q_a = 1$, and $q_b = -0.5$ and identical polarization inputs (HH/VV). The evolution from $\delta_b = 0$ to $\pi$ illustrates the transition from single-beam dominated patterns to hybrid topological structures. \textbf{(b)} Visibility distributions ($\nu$) for two distinct regimes: $\delta_a = \pi/6$ (left) and $\delta_a = 5\pi/6$ (right). The vertical progression shows the shift from broad interference fringes to high-frequency or homogeneous distributions. Note that $q_a=1$ and $q_b=-0.5$. The terminal line charts quantify the weights of the fundamental modes $\Delta_i(\delta)$ as a function of $\delta_b$. These weights determine the proportional contribution of each geometric mode to the final interference pattern, highlighting the role of $\delta_a$ as a macroscopic switch for selecting the system's asymptotic state.} 
\label{fig:change_delta}
\end{figure}

Figure \ref{fig:berryphase}\textbf{(b)} illustrates the effects of varying $q_b$ from $0.5$ to $20$ while holding $q_a = 1, \delta_a=\pi/2, \delta_b=\pi/2$ constant. We observe that topological charge scales the spatial frequency of the interference pattern; for $\ket{\phi^+}$, fringes arise when $q_a \neq q_b$, as per Eq. \eqref{eq:bpx}, and disappear when charges are identical. We observe that increasing $q_b$ increases the density of phase cycles within the $[-\pi, \pi]$ coordinate space, thereby increasing the number of phase singularities and {pathways} with topological invariance. {This invariance is defined by the conservation of the singularity count within the coincidence manifold, which remains constant under local perturbations such as detuning. Consequently, the monotonic scaling of these defects with $q_b$ establishes a robust mapping between the topological signature and the state’s dimensionality.} Despite these structural differences, the BPX evolution remains consistent across all four Bell states. Notably, phase singularities persist even at small detunings or low topological charges. These BPX characteristics infer the resulting spatial patterning of Bell state probabilities and interference visibility, shown in Figs. \ref{fig:change_delta} and \ref{fig:change_q}, respectively.

\begin{figure}
\includegraphics [width=\columnwidth]{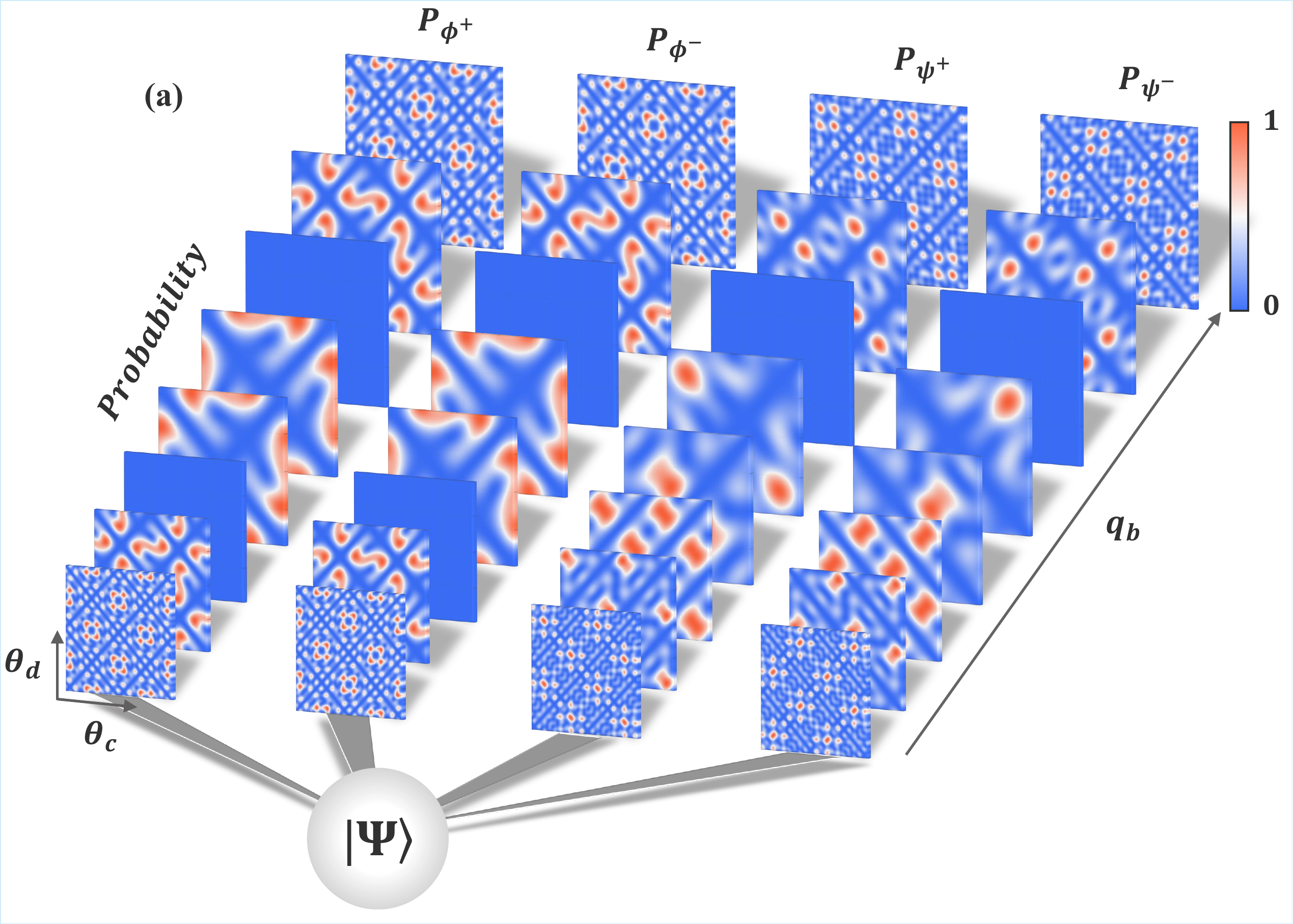}\\ [5mm]
\includegraphics [width=1.15\columnwidth]{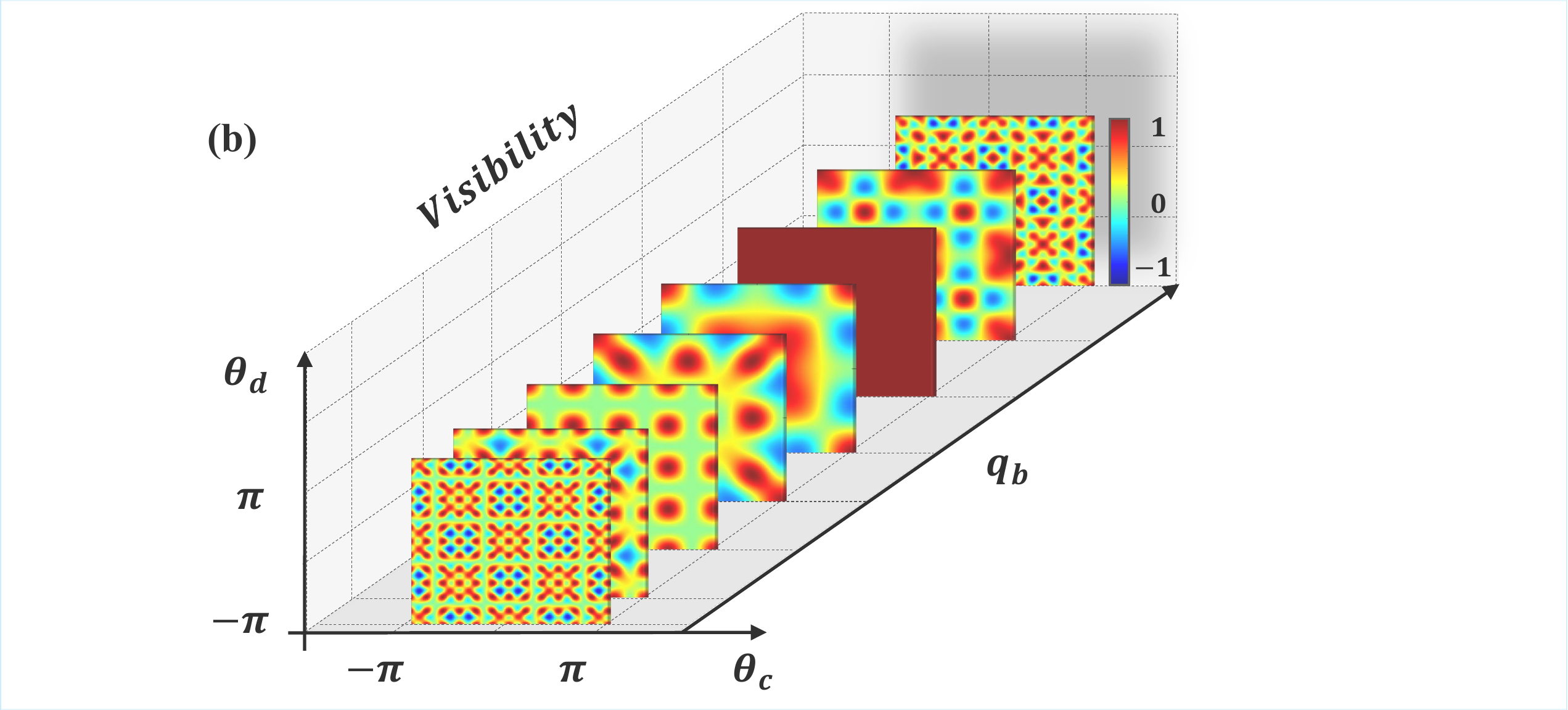}
\centering
\caption{\textbf{Topological scaling and symmetry of spatial correlations.} \textbf{(a)} Normalized spatial probability distributions of the four Bell states as a function of path-$b$ $q$-plate topological charge, $q_b$. The evolution from $q_b = -5$ to $q_b=5$ illustrates the multiplicative effect of the charge on the spatial frequency of interference fringes. The overall impinging pattern is a direct consequence of the patterning dictated by BPX. \textbf{(b)} Interference visibility spatial distributions across varying topological charges. The vertical progression illustrates the transition from high-frequency, complex lattice-like interference fringes for large negative $q_b$ to broad spatial features at small $|q_b|$, then returning to the high-frequency features. All subplots are calculated for $q_a=1$, $\delta_a=\delta_b=\pi/2$, and identical polarization (HH/VV) inputs, mapped over the $(\theta_c, \theta_d)$ azimuthal coordinate space. The results demonstrate that while the spatial frequency is driven by the topological charge $q_b$, the fundamental symmetry and normalization rules of the Bell basis remain topologically protected. }
\label{fig:change_q}
\end{figure}

{Note that Bell state probabilities and the visibility term can be expressed in the unified form:
\begin{equation}
\sum_{i=1}^{5} Q_i(q) \cdot \Delta_i(\delta) \cdot \Theta_i(\theta).
\end{equation}
Spatial distributions are treated as a superposition of five fundamental modes, $\Theta_i(\theta)$, weighted by coefficients $Q_i(q)$ and $\Delta_i(\delta)$. The shared nature of these coefficients ensures both probability and visibility follow the same variation rules across fixed mode pairings (See Supplementary Material for details).}

With HH/VV inputs, Fig. \ref{fig:change_delta}\textbf{(a)} shows the evolution of the four Bell states initially dominated by the path-$a$ $q$-plate into a hybrid ($q_a,q_b$)-type structure as $\delta_b$ increases. We note that $q_a=1, q_b=-0.5$ and $\delta_a=\pi/2$. Isolated spots transform into continuous diagonal fringes, reflecting the strong modulatory effect of $\delta_b$ on the system’s spatial profile. Figure \ref{fig:change_delta}\textbf{(b)} shows the visibility ($\nu$) progression for two distinct regimes: $\delta_a = \pi/6$ and $\delta_a = 5\pi/6$. For $\delta_a = \pi/6$ (left), the visibility is near unity, indicating high indistinguishability presenting as a HOM peak for small $\delta_b$ with $q_a=1$ and $q_b=-0.5$.  As $\delta_b$ increases, clear interference fringes arise with distinct regions of HOM dips (blue) and peaks (red), as well as distinguishable regions (green). Conversely, at $\delta_a = 5\pi/6$ (right), the system maintains a high-spatial-frequency grid structure with slight modifications to the interference fringe density. We find that $\delta_a$ acts as a macroscopic switch for the system's final state, indicated by the visibility components $\Delta_i$. We observe that $\Delta_5$ ($\Delta_1$) transitions to $\Delta_2$ ($\Delta_3$) as $\delta_b$ increases for  $\delta_a=\pi/6$ ($\delta_a=5\pi/6$), which effectively enables deterministic control of the system’s final state. {Overall, detuning chooses which path's q-plate spatial pattern dominates, while $q$ controls the specific spatial morphology, such as stripes, and spatial frequency.}


Figure \ref{fig:change_q} illustrates the influence of topological charge $q_b$ on the spatial patterns. The evolution of the four Bell state probabilities, $P_{\phi^\pm}$ and $P_{\psi^\pm}$, is plotted as a function of $q_b \in [-5, 5]$. For large topological charges, we observe complex, high-spatial-frequency lattice fringes with strong contrast. As $q_b$ approaches zero, these features transition into broader, smoother structures. While the states share a consistent evolutionary trajectory along the $q_b$ axis, their specific spatial features, such as peak positions, nodal lines, and symmetries, differ fundamentally at any given cross-section. However, the topological invariance within each plane remains intact, offering a set of design parameters to produce noise-robust quantum states. Following a parallel-like progression, the interference visibility is plotted in Fig. \ref{fig:change_q}\textbf{(b)}, transitioning from a dense lattice at $q_b = \pm5$ to lower-frequency patterns near $q_b = 0$. 

Collectively, we find that Bell state probability spatial distributions are the same for identical polarization inputs (HH or VV). Orthogonal inputs (HV and VH) produce similar symmetry in behaviour. As for visibility, $C_{\mathrm{in}}$ also exhibits this same polarization dependence; $C_{\mathrm{out}}$ remains invariant across all input combinations. In this case, two distinct sets of inputs can generate vastly different quantum states with unique properties and robustness, useful for a variety of quantum applications.


\section*{Conclusions}
In conclusion, we have established a precise mathematical framework for  HOM interference of topological quantum states under arbitrary $q$-plate detunings and topological charges. We demonstrated that Bell state probability distributions and interference visibilities can be elegantly decomposed into a superposition of fundamental modes governed by geometry, $\Theta_i(\theta)$. This behavior is led by  BPX, an interaction geometric phase that dictates the dynamic evolution, such as transitions between symmetric and antisymmetric manifolds based on macroscopic control parameters. Furthermore, by monitoring the BPX, we identified a non-tomographic witness that enables the estimation of state dimensionality through topological invariants and phase singularities. This approach allows the Hilbert space dimension to be inferred without full-state tomography, significantly reducing the computational overhead for high-dimensional characterization. 
BPX generalization to multipartite systems is also possible, allowing further understanding of the geometric dependence of Hilbert space expansion by increasing particle number.
Our results provide a rigorous foundation for understanding structured light interference under non-ideal conditions, offering a pathway toward robust state preparation~\cite{bauer2015observation},  the development of high-capacity quantum communication networks~\cite{sit2017high}, and enhanced sensitivity in topological quantum metrology~\cite{brasselet2018tunable}. {Moreover, the potential non-Abelian nature of the system provides insight toward applications in non-Abelian physics and non-Abelian holonomic quantum computing~\cite{zhang2022non, pachos1999non}.}

\section*{Acknowledgments}
This work was supported by the National Natural Science Foundation of China (Nos. 12574392, W2533027, U25A20194, 12334015, 92463308); the Natural Science Foundation of Jiangsu Province (Nos. BK20233001, BK20252118); and the Jiangsu Province Excellent Postdoctoral Fellowship Program (No. 2025ZB498).

\bibliography{refs}

\clearpage

\onecolumngrid

\section*{Supplementary Material for: Topological Quantum Interferometry}

\section*{S1. THEORETICAL DERIVATION}
The effect of the q-plate acting on the polarization basis $\{\ket{H},\ket{V}\}$, where $\ket{L}=\ket{H}-i\ket{V}$ and $\ket{R}=\ket{H}+i\ket{V}$, is described as
\begin{equation}
\hat{U}_{q\text{-}plate}=\hat{R}(-\gamma)\hat{D}(\delta)\hat{R}(\gamma),
\end{equation}
which comprises a rotation matrix, $\hat{R}(\pm\gamma)$ and a phase matrix $\hat{D}(\delta)$, where
\begin{equation}
    \hat{R}(\gamma) = \begin{pmatrix}
    \cos(\gamma) & \sin(\gamma) \\
   -\sin(\gamma) & \cos(\gamma)
    \end{pmatrix}
\end{equation}
and 
\begin{equation}
    \hat{D}(\delta) =\begin{pmatrix}
    \exp(i\delta/2) & 0 \\
   0 & \exp(-i\delta/2)
    \end{pmatrix}.
\end{equation}
Here, $\gamma=2q\theta$, $q$ represents the topological charge and $\theta$ represents the azimuthal coordinate. The constituents are scaled by radially dependent hypergeometric functions, $F_q (r)$ ($F_0(r)$ for $q=0$) written as~\cite{karimi2007hypergeometric,allen1992orbital,gao2025generation}:
\begin{equation}\label{eq:2}
F_q\left(r\right)=\sqrt{\frac{1}{\pi\left(|q|\right)!}}\times\frac{1}{w_0}\times{\left(\frac{r\sqrt2}{w_0}\right)}^{|2q|}\times e^{-\frac{r^2}{{w_0}^2}}
\end{equation}
where $w_{0}$ is the beam waist.

We model quantum states of two incident light beams passing through an optical path where the polarization states are set by quarter- and half-wave plates 
resulting in four polarization combinations: HH, HV, VH, and VV. These two modulated polarized beams pass through the q-plate and subsequently enter a BS to interfere.
Taking the incident polarization combination HH as an example, the quantum state of the two photons after passing through the q-plate can be calculated as
\begin{equation}\label{eq:3}
\begin{aligned}
\hat{U}^{a}_{\mathrm{q\text{-}plate}}\hat{U}^{b}_{\mathrm{q\text{-}plate}}\cdot \ket{H}_{a}\ket{H}_{b}=
    &\left\{F_{0}(r_a)\cos{\frac{\delta_a}{2}}\ket{H}_{a}+i\sin{\frac{\delta_a}{2}}F_{q_a}(r_a)(\cos{2q_{a}\theta_{a}}\ket{H}_{a}+\sin{2q_{a}\theta_{a}}\ket{V}_{a})\right\}  \\
    &\left\{F_{0}(r_b)\cos{\frac{\delta_b}{2}}\ket{H}_{b}+i\sin{\frac{\delta_b}{2}}F_{q_b}(r_b)(\cos{2q_{b}\theta_{b}}\ket{H}_{b}+\sin{2q_{b}\theta_{b}}\ket{V}_{b})\right\}  \\
\end{aligned}
\end{equation}
where sub-(super-)scripts $a$ and $b$ represent the paths of the photons.

The beams then enter a BS, which has an action  written as
\begin{equation}
    \hat{\mathcal{BS}} =\frac{1}{\sqrt{2}}\begin{pmatrix}
    1 & 1 \\
   1 & -1
    \end{pmatrix}.
\end{equation}
The BS acts on initial states $\Phi_{a}$ $(\Phi_b)$ producing a sum (difference) of output states $\Phi_c$ and $\Phi_d$.

From here, we make the following simplification: all other output modes except when two output ports emit exactly one photon are disregarded. By invoking the BS action on Eq.~\eqref{eq:3} and simplifying, we obtain the final state, $\ket{\Psi}$, as
\begin{equation}\label{eq:5}
\ket{\Psi}=\mathcal{N}\left(c_{HH}\ket{H}_{c}\ket{H}_{d}+c_{VV}\ket{V}_{c}\ket{V}_{d}+c_{HV}\ket{H}_{c}\ket{V}_{d}+c_{VH}\ket{V}_{c}\ket{H}_{d}\right)
\end{equation}
where $\mathcal{N}$ is the normalization factor, subscripts $c$ and $d$ denote the photon output ports, and $c_{ij}$ are the coefficients for each output polarization combination, with $i,j$ taking $H$ or $V$. Equation \eqref{eq:5} can now be rewritten in the Bell basis as follows:
\begin{equation}\label{eq:6}
\ket{\Psi}=\sum_{\mathcal{B}} c_{\mathcal{B}} \ket{\mathcal{B}}=\mathcal{N}\left(c_{\phi^{+}}\ket{\phi^{+}}+c_{\phi^{-}}\ket{\phi^{-}}+c_{\psi^{+}}\ket{\psi^{+}}+c_{\psi^{-}}\ket{\psi^{-}}\right),
\end{equation}
where $\mathcal{B} \in {\psi^{\pm}, \phi^{\pm}}$.
Here, the corresponding relationships between the coefficients are $c_{\phi^{\pm}}=A_c\left(c_{HH}\pm c_{VV}\right)$ and $c_{\psi^{\pm}}=A_c\left(c_{HV}\pm c_{VH}\right)$ with $A_c=1/\sqrt{2}$.

When the impact of $F_q$ needs to be considered, the function can be integrated over the entire space to obtain a normalization factor. In general, there are three important contributors when considering the interaction term amplitudes. Specifically, we look at the results of 
$F_0^2F_{a}^2$ or $F_0^2F_{b}^2$, $F_a^2F_b^2$, and $F_0^2F_aF_b$. The results of these three terms under integration are as follows
\begin{equation}
\begin{aligned}
   &\int_{0}^{\infty} F_{0}^{2}F_{a,b}^{2} rdr = \frac{1}{8\pi^{2}} \frac{1}{{4}^{\vert q_{a,b} \vert}} \frac{1}{\vert q_{a,b}\vert!} \Gamma(2\vert q_{a,b} \vert + 1)  \\
   &\int_{0}^{\infty} F_{a}^{2}F_{b}^{2} \, rdr = \frac{1}{8\pi^{2}} \frac{1}{{4}^{\vert q_{a} \vert + \vert q_{b} \vert}} \frac{1}{\vert q_{a} \vert! \vert q_{b} \vert! } \Gamma(2\vert q_{a} \vert + 2\vert q_{b} \vert + 1)  \\
   &
   \int_{0}^{\infty} F_{0}^{2}F_{a}F_{b} \, rdr = \frac{1}{8\pi^{2}} \frac{1}{{2}^{\vert q_{a} \vert + \vert q_{b} \vert}} \frac{1}{\sqrt{\left|q_a\right|! \left|q_b\right|!}} \Gamma(\vert q_{a} \vert + \vert q_{b} \vert + 1)  \\
   \end{aligned}.
\end{equation}
However, to better understand the azimuthal dependence with respect to the detuning and topological, we set $F_q$ and $F_0$ equal to  1 in the main text. 
\section*{S2. CALCULATION OF FOUR BELL STATE PROBABILITY AND INTERFERENCE VISIBILITY}

The probability of four Bell states can be calculated by taking the square of $c$-coefficient terms in \eqref{eq:6}, that is $P_{\mathcal{B}}=|c_{\mathcal{B}}|^2$. 
For the calculation of visibility, we  follow the form of Ref.~\cite{schiano2026tailoring}, written as $\nu=\left(C_{\mathrm{out}}-C_{\mathrm{in}}\right)/C_{\mathrm{out}}$. 
The probability $P_{\mathcal{B}}$ and visibility $\nu$ share a common functional form as seen in the main text, $\sum_{i=1}^{5}{Q_i\left(q\right)\cdot\Delta_i\left(\delta\right)\cdot\Theta_i\left(\theta\right)}$. 

For the functions $Q_i\left(q\right)$ and $\Delta_i\left(\delta\right)$, they have fixed pairings, as detailed in Table~\ref{tab:1}.
\begin{table}[H]
\centering
\caption{The fixed pairs of $Q_i\left(q\right)$ and $\Delta_i\left(\delta\right)$}
\begin{tabular}{|c|c|c|c|c|c|}
    \hline
    $i$ & 1 & 2 & 3 & 4 & 5 \\
    \hline
    $Q_i\left(q\right)$ & $F_0^2 F_a^2$ & $F_0^2 F_b^2$ & $F_a^2 F_b^2$ & $F_0^2 F_a F_b$ & $F_0^4$\\
    \hline
    $\Delta_i\left(\delta\right)$ & $\sin^2{\frac{\delta_a}{2}}\cos^2{\frac{\delta_b}{2}}$ & $\cos^2{\frac{\delta_a}{2}}\sin^2{\frac{\delta_b}{2}}$ & $\sin^2{\frac{\delta_a}{2}}\sin^2{\frac{\delta_b}{2}}$ & $\frac{1}{4}\sin{\delta_a}\sin{\delta_b}$ & $\cos^2{\frac{\delta_a}{2}}\cos^2{\frac{\delta_b}{2}}$\\
    \hline
\end{tabular}
\label{tab:1}
\end{table}

Here, $F_a$ and $F_b$ in the table are abbreviations for $F_{q_a}$ and $F_{q_b}$. While the specific form of $\Theta_i(\theta)$ depends on both the input state and the quantity being calculated.
We provide the individual calculation results below. 
For the Bell state probabilities when the input state is HH or VV, the calculation result for $\Theta_i(\theta)$ is given by Table~\ref{tab:2}.
As for when the input state is HV or VH, the calculation result for $\Theta_i(\theta)$ is given by Table~\ref{tab:3}.
\begin{table}[H]
\centering
\caption{Specific forms of functions $\Theta_i(\theta)$ in the Bell state distribution for HH/VV input states}
\begin{tabular}{|c|c|c|}
\hline
$i$                  & Bell states & $\Theta_i(\theta)$                                                                         \\ \hline
\multirow{4}{*}{$1$} & $\phi^+$    & $(\cos 2q_a\theta_c - \cos 2q_a\theta_d)^2$                                                \\ \cline{2-3} 
                     & $\phi^-$    & Same as $\phi^+$                                                                           \\ \cline{2-3} 
                     & $\psi^+$    & $(\sin 2q_a\theta_c - \sin 2q_a\theta_d)^2$                                                \\ \cline{2-3} 
                     & $\psi^-$    & $(\sin 2q_a\theta_c + \sin 2q_a\theta_d)^2$                                                \\ \hline
\multirow{4}{*}{$2$} & $\phi^+$    & $(\cos 2q_b\theta_c - \cos 2q_b\theta_d)^2$                                                \\ \cline{2-3} 
                     & $\phi^-$    & Same as $\phi^+$                                                                           \\ \cline{2-3} 
                     & $\psi^+$    & $(\sin 2q_b\theta_c - \sin 2q_b\theta_d)^2$                                                \\ \cline{2-3} 
                     & $\psi^-$    & $(\sin 2q_b\theta_c + \sin 2q_b\theta_d)^2$                                                \\ \hline
\multirow{4}{*}{$3$} & $\phi^+$    & $(\cos(2q_a\theta_c - 2q_b\theta_d) - \cos(2q_a\theta_d - 2q_b\theta_c))^2$                \\ \cline{2-3} 
                     & $\phi^-$    & $(\cos(2q_a\theta_c + 2q_b\theta_d) - \cos(2q_a\theta_d + 2q_b\theta_c))^2$                \\ \cline{2-3} 
                     & $\psi^+$    & $(\sin(2q_a\theta_c + 2q_b\theta_d) - \sin(2q_a\theta_d + 2q_b\theta_c))^2$                \\ \cline{2-3} 
                     & $\psi^-$    & $(\sin(2q_a\theta_c - 2q_b\theta_d) + \sin(2q_a\theta_d - 2q_b\theta_c))^2$                \\ \hline
\multirow{4}{*}{$4$} & $\phi^+$    & $-2(\cos 2q_a\theta_c - \cos 2q_a\theta_d)(\cos 2q_b\theta_c - \cos 2q_b\theta_d)$         \\ \cline{2-3} 
                     & $\phi^-$    & Same as $\phi^+$                                                                           \\ \cline{2-3} 
                     & $\psi^+$    & $-2(\sin 2q_a\theta_c - \sin 2q_a\theta_d)(\sin 2q_b\theta_c - \sin 2q_b\theta_d)$         \\ \cline{2-3} 
                     & $\psi^-$    & $-2(\sin 2q_a\theta_c + \sin 2q_a\theta_d)(\sin 2q_b\theta_c + \sin 2q_b\theta_d)$         \\ \hline
\multirow{4}{*}{$5$} & $\phi^+$    & $0$                                                                                        \\ \cline{2-3} 
                     & $\phi^-$    & $0$                                                                                        \\ \cline{2-3} 
                     & $\psi^+$    & $0$                                                                                        \\ \cline{2-3} 
                     & $\psi^-$    & $0$                                                                                        \\ \hline
\end{tabular}
\label{tab:2}
\end{table}

\newpage

\begin{table}[H]
\centering
\caption{Specific forms of functions $\Theta_i(\theta)$ in the Bell state distribution for HV/VH input states.}
\begin{tabular}{|c|c|c|}
\hline
$i$ & Bell states & $\Theta_i(\theta)$ \\ \hline
\multirow{4}{*}{$1$} & $\phi^+$ & $(\sin 2q_a\theta_c - \sin 2q_a\theta_d)^2$ \\ \cline{2-3} 
 & $\phi^-$ & Same as $\phi^+$ \\ \cline{2-3} 
 & $\psi^+$ & $(\cos 2q_a\theta_c - \cos 2q_a\theta_d)^2$ \\ \cline{2-3} 
 & $\psi^-$ & $(\cos 2q_a\theta_c + \cos 2q_a\theta_d)^2$ \\ \hline
\multirow{4}{*}{$2$} & $\phi^+$ & $(\sin 2q_b\theta_c - \sin 2q_b\theta_d)^2$ \\ \cline{2-3} 
 & $\phi^-$ & Same as $\phi^+$ \\ \cline{2-3} 
 & $\psi^+$ & $(\cos 2q_b\theta_c - \cos 2q_b\theta_d)^2$ \\ \cline{2-3} 
 & $\psi^-$ & $(\cos 2q_b\theta_c + \cos 2q_b\theta_d)^2$ \\ \hline
\multirow{4}{*}{$3$} & $\phi^+$ & $(\sin(2q_a\theta_c - 2q_b\theta_d) - \sin(2q_a\theta_d - 2q_b\theta_c))^2$ \\ \cline{2-3} 
 & $\phi^-$ & $(\sin(2q_a\theta_c + 2q_b\theta_d) - \sin(2q_a\theta_d + 2q_b\theta_c))^2$ \\ \cline{2-3} 
 & $\psi^+$ & $(\cos(2q_a\theta_c + 2q_b\theta_d) - \cos(2q_a\theta_d + 2q_b\theta_c))^2$ \\ \cline{2-3} 
 & $\psi^-$ & $(\cos(2q_a\theta_c - 2q_b\theta_d) + \cos(2q_a\theta_d - 2q_b\theta_c))^2$ \\ \hline
\multirow{4}{*}{$4$} & $\phi^+$ & $-2(\sin 2q_a\theta_c - \sin 2q_a\theta_d)(\sin 2q_b\theta_c - \sin 2q_b\theta_d)$ \\ \cline{2-3} 
 & $\phi^-$ & $2(\sin 2q_a\theta_c - \sin 2q_a\theta_d)(\sin 2q_b\theta_c - \sin 2q_b\theta_d)$ \\ \cline{2-3} 
 & $\psi^+$ & $2(\cos 2q_a\theta_c - \cos 2q_a\theta_d)(\cos 2q_b\theta_c - \cos 2q_b\theta_d)$ \\ \cline{2-3} 
 & $\psi^-$ & \begin{tabular}[c]{@{}l@{}}$-2(\cos 2q_a\theta_c - \cos 2q_a\theta_d)(\cos 2q_b\theta_c - \cos 2q_b\theta_d)$ \\ $+ 4(\sin 2q_a\theta_c \sin 2q_b\theta_d + \sin 2q_a\theta_d \sin 2q_b\theta_c)$\end{tabular} \\ \hline
\multirow{4}{*}{$5$} & $\phi^+$    & $0$                                                                                        \\ \cline{2-3} 
                     & $\phi^-$    & $0$                                                                                        \\ \cline{2-3} 
                     & $\psi^+$    & $0$                                                                                        \\ \cline{2-3} 
                     & $\psi^-$    & $4$                                                                                        \\ \hline
\end{tabular}
\label{tab:3}
\end{table}

\begin{table}[H]
\centering
\caption{Specific forms of functions $\Theta_i(\theta)$ in the $C_{\mathrm{out}}-C_{\mathrm{in}}$ term of the visibility for HH/VV input states.}
\begin{tabular}{|c|c|}
\hline
$i$& $\Theta_i(\theta)$ \\ \hline
$1$ & $\cos(2q_a\theta_c)\cos(2q_a\theta_d)$ \\ \hline
$2$ & $\cos(2q_b\theta_c)\cos(2q_b\theta_d)$ \\ \hline
$3$ & $\cos(2(q_a-q_b)\theta_c)\cos(2(q_a-q_b)\theta_d)$ \\ \hline
$4$ & $(\cos 2q_a\theta_c-\cos 2q_a\theta_d)(\cos 2q_b\theta_c-\cos 2q_b\theta_d) + \sin 2q_a\theta_c \sin 2q_b\theta_c + \sin 2q_a\theta_d \sin 2q_b\theta_d$ \\ \hline
$5$ & 1 \\ \hline
\end{tabular}
\label{tab:4}
\end{table}


\begin{table}[H]
\centering
\caption{Specific forms of functions $\Theta_i(\theta)$ in the $C_{\mathrm{out}}-C_{\mathrm{in}}$ term of the visibility for HV/VH input states.}
\begin{tabular}{|c|c|}
\hline
$i$& $\Theta_i(\theta)$ \\ \hline
$1$ & $\sin(2q_a\theta_c)\sin(2q_a\theta_d)$ \\ \hline
$2$ & $\sin(2q_b\theta_c)\sin(2q_b\theta_d)$ \\ \hline
$3$ & $\sin(2(q_a-q_b)\theta_c)\sin(2(q_a-q_b)\theta_d)$ \\ \hline
$4$ & $-\sin(2q_a\theta_c)\sin(2q_b\theta_d) - \sin(2q_b\theta_c)\sin(2q_a\theta_d)$ \\ \hline
$5$ & 0 \\ \hline
\end{tabular}
\label{tab:5}
\end{table}

The calculation of the visibility relies on examining the dependencies of both the numerator $C_{\mathrm{out}}-C_{\mathrm{in}}$ and the denominator $C_{\mathrm{out}}$. In Table \ref{tab:4}, we calculate $C_{\mathrm{out}}-C_{\mathrm{in}}$ for an input state that is HH or VV, and represented by $\Theta_i(\theta)$. 
When the input state is HV or VH, the calculation result is given by Table~\ref{tab:5}. As for $C_{out}$, this term is independent of $\theta$ and remains invariant under any input states (HH, VV, HV, and VH). The specific term is given by Table~\ref{tab:6}.

\begin{table}[H]
\centering
\caption{Specific forms of functions $\Theta_i(\theta)$ in the $C_{out}$ term of the visibility for HH/HV/VH/VV input states.}
\begin{tabular}{|c|c|}
\hline
$i$& $\Theta_i(\theta)$ \\ \hline
$1$ & 2 \\ \hline
$2$ & 2 \\ \hline
$3$ & 2 \\ \hline
$4$ & 0 \\ \hline
$5$ & 2 \\ \hline
\end{tabular}
\label{tab:6}
\end{table}

\section*{S3. Spatial dependence of Bell state probability and interference visibility for orthogonal inputs}

In Figs.~\ref{fig:change_q}-\ref{fig:change_delta}, we provide additional results for orthogonal polarization inputs (HV/VH) with the same conditions as in the main text.

\begin{figure*}[tbph]
\includegraphics [width=0.45\columnwidth]{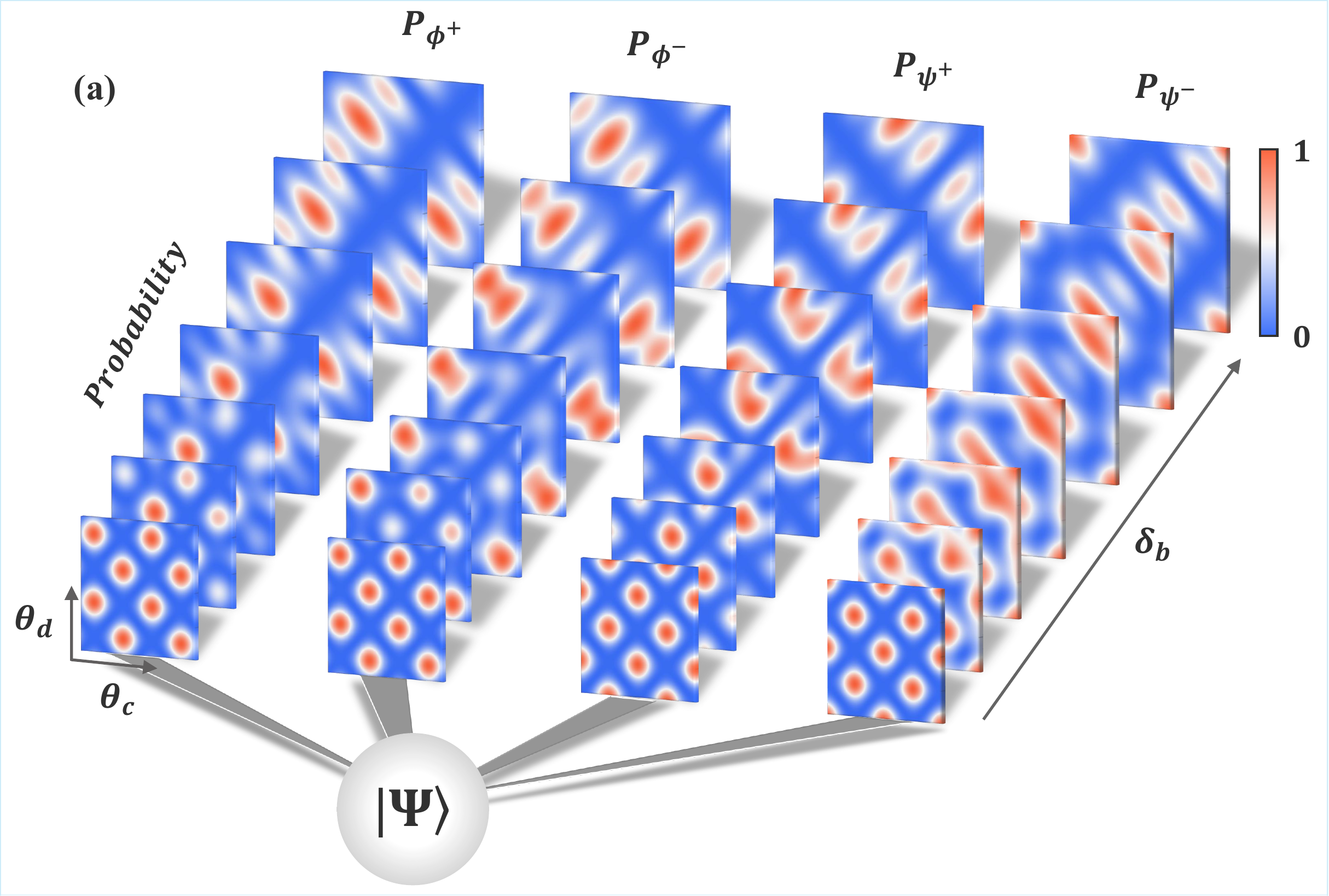}
\includegraphics [width=0.45\columnwidth]{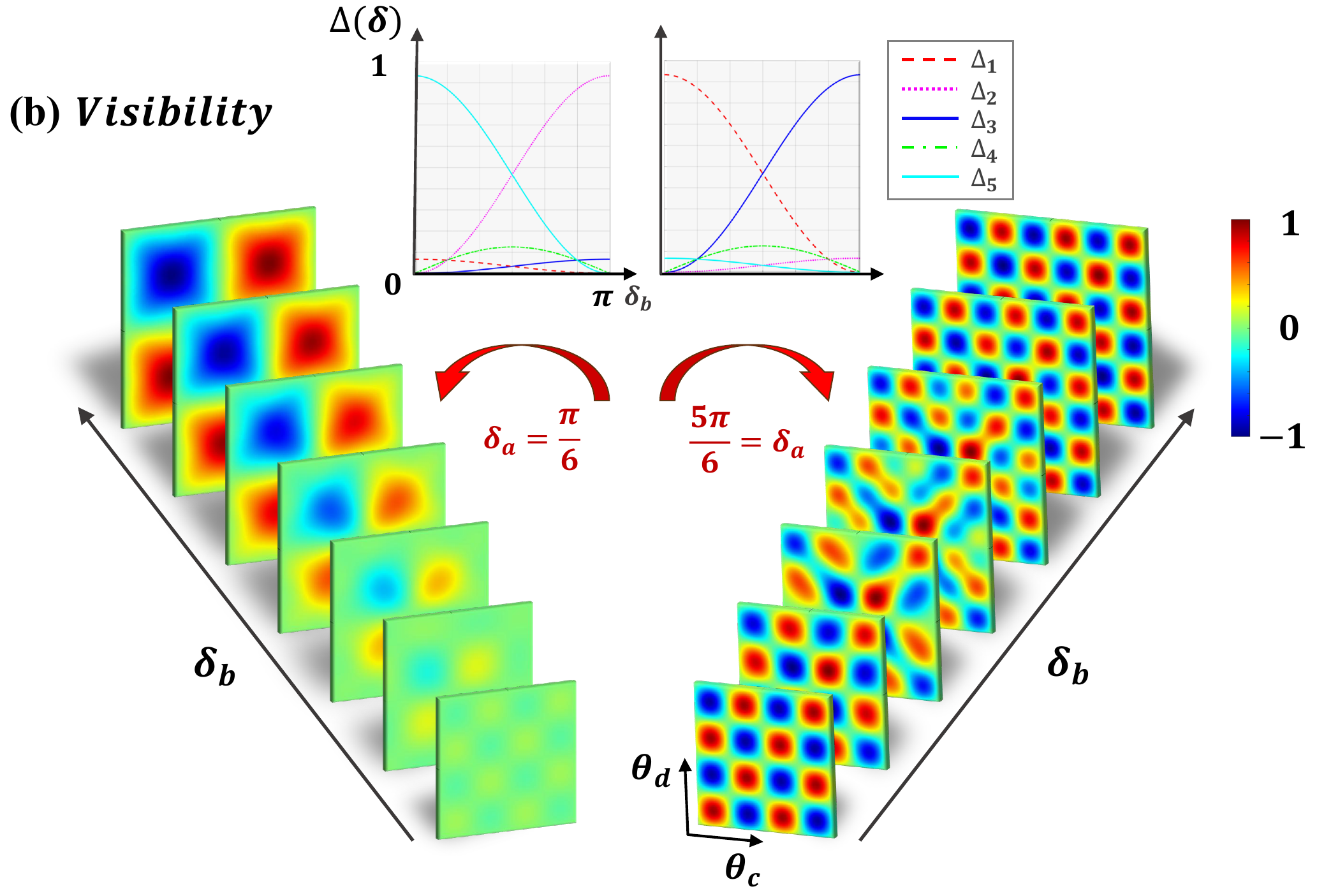}
\centering
\caption{\textbf{Spatial correlations for various topological charges for orthogonal input polarizations.} \textbf{(a)} Subplots detailing the spatial probability distributions of the Bell states. The four columns represent the respective distributions for each of the four Bell states. The results have been normalized. \textbf{(b)} Subplots illustrating the visibility distributions. The data in both panels is generated using HV/VH input light and plotted over the $(\theta_c, \theta_d) \in [-\pi, \pi]$ space, keeping $q_a=1$ and $\delta_a=\delta_b=\pi/2$ constant. The vertical progression highlights the effect of varying $q_b$, evaluated at $q_b \in \{-5, -2, -1, -0.5, 0.5, 1, 2, 5\}$.}
\label{fig:change_q}
\end{figure*}

\begin{figure*}[tbph]
\includegraphics [width=0.4\columnwidth]{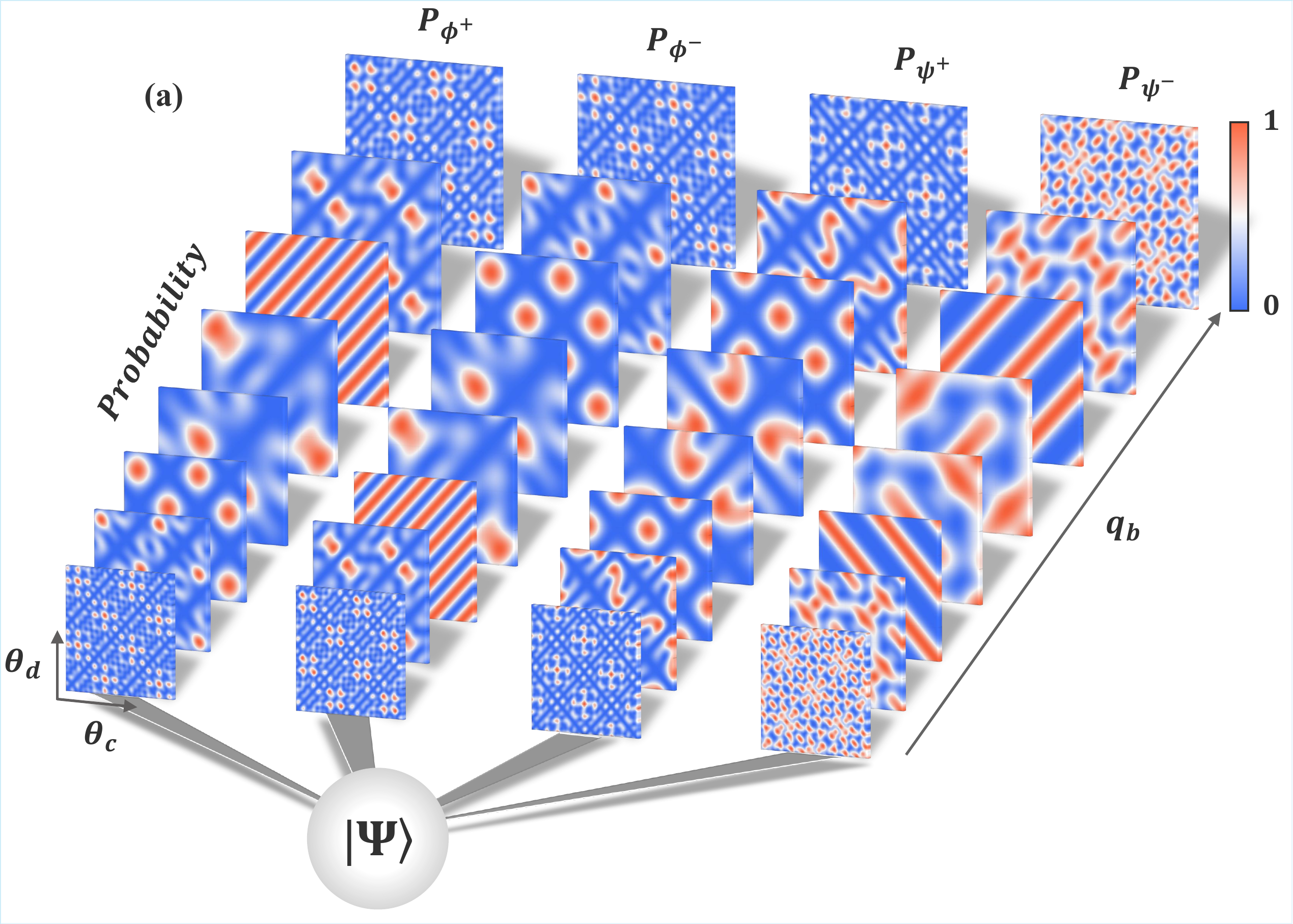}
\includegraphics [width=0.5\columnwidth]{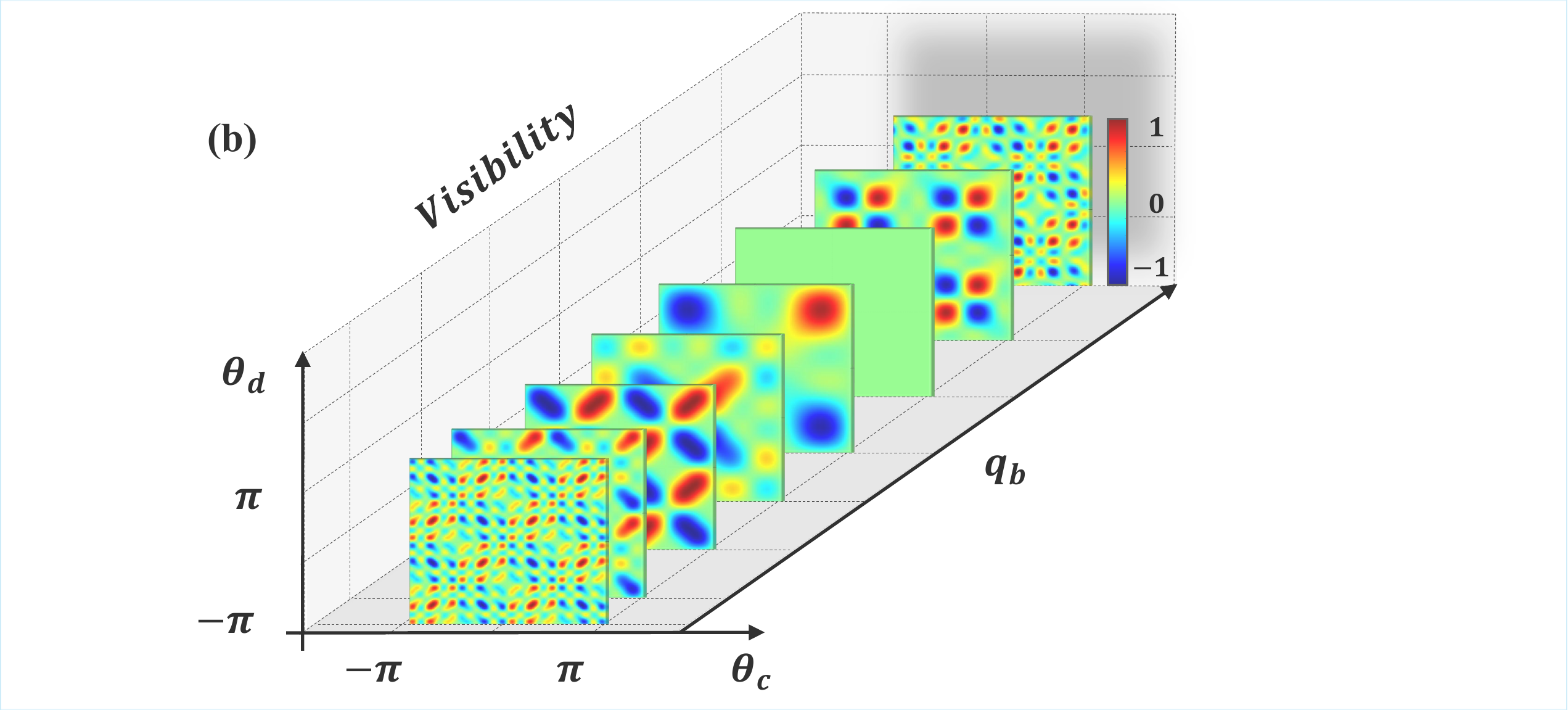}
\centering
\caption{\textbf{Spatial patterns under various detuning for orthogonal input polarizations} \textbf{(a)} Spatial probability distributions of the Bell states for $\delta_a=\pi/2$. The four columns represent the respective distributions for each of the four Bell states. The results have been normalized. \textbf{(b)} Spatial visibility distributions for $\delta_a=\pi/6$ (left path) and $\delta_a=5\pi/6$ (right path). The line charts at the end of each path illustrate the variation of the $\Delta(\delta)$ functions with respect to $\delta_b$, which determines the proportion of each mode in the interference pattern. All subplots in (a) and (b) are plotted over the $(\theta_c, \theta_d)$ parameter space within the interval $[-\pi, \pi]$ for an HV/VH input state, under the fixed conditions of $q_a=1$ and $q_b=-0.5$. The vertical arrangement demonstrates the evolution of these distributions as a function of $\delta_b$, where $\delta_b \in \{0, \frac{\pi}{6}, \frac{\pi}{3}, \frac{\pi}{2}, \frac{2\pi}{3}, \frac{5\pi}{6}, \pi\}$.}
\label{fig:change_delta}
\end{figure*}

\end{document}